%% file: A4_Constellation_Final.tex
\documentclass[conference]{IEEEtran}
\usepackage{algorithm,algorithmic}
\usepackage{amsmath,amssymb,mathrsfs}
\usepackage{array}
\usepackage{balance}
\usepackage{booktabs}
\usepackage{cite}
\usepackage{epsfig}
\usepackage{fancyhdr}
\usepackage{float}
\usepackage[nomain,acronym,automake,toc,nopostdot]{glossaries}
\usepackage{graphicx,color}
\usepackage{mathtools}
\usepackage{multirow}
\usepackage{pgfplots}
\usepackage{psfrag}
\usepackage{stfloats}
\usepackage{subfigure}

\newtheorem{lem}{Lemma}

\newtheorem{thm}{Theorem}

\makeglossaries
\input{Acronyms}

\definecolor{sblue}{RGB}{0,51,120}
\definecolor{sred}{RGB}{200,51,130}

\newcommand{\figref}[1]{Fig. \ref{#1}}

\newcommand{\thmref}[1]{{\it Theorem \ref{#1}}}

\newcommand{\lemref}[1]{{\it Lemma \ref{#1}}}
\newcommand{\appref}[1]{{\textsc{Appendix} \ref{#1}}}
\renewcommand{\eqref}[1]{(\ref{#1})}

\ifCLASSINFOpdf
\else
\fi

\begin{document}
\title{LO-Aware Adaptive Modulation for \\ Rydberg Atomic Receivers}
\author{Jiuyu Liu, Yi Ma$^\dagger$, and Rahim Tafazolli\\{\small Institute for Communication Systems (ICS), University of Surrey, Guildford, U.K.}\\{\small E-mail: (jiuyu.liu, y.ma, r.tafazolli)@surrey.ac.uk}}
{}

\maketitle

\begin{abstract}
	Rydberg atomic (RA) receivers represent a revolutionary quantum technology for wireless communications, offering unprecedented sensitivity beyond conventional radio frequency (RF) antennas.
	However, these receivers detect only signal amplitude, losing critical phase information.
	While reference signals generated by a local oscillator (LO) can assist in phase recovery, existing modulation schemes designed for conventional systems perform poorly with this quantum detection mechanism.
	This paper introduces a breakthrough LO-aware adaptive modulation (LOAM) scheme specifically developed for RA receivers that dynamically adapts to complex fading channel coefficients.
	LOAM maximizes the minimum amplitude difference between constellation points, ensuring optimal detection performance.
	The innovation employs an adaptive co-linear constellation architecture aligned with the combined phase of reference signal and channel coefficient.
	For strong reference signals, LOAM generates symmetric constellation points centered at origin; for weak signals, it adopts non-symmetric distributions.
	The paper mathematically derives the threshold governing these operational regimes.
	Simulation results reveal the transformative impact of LOAM, demonstrating performance gains exceeding 45 dB over conventional modulation schemes, including quadrature amplitude modulation (QAM), phase-shift keying (PSK), and pulse-amplitude modulation (PAM).
\end{abstract}


\section{Introduction}\label{sec1}
Rydberg atomic (RA) receivers, originally developed for quantum sensing, have emerged as a transformative technology for wireless communications \cite{Fancher2021}.
These quantum devices surpass conventional radio frequency (RF) antennas across multiple dimensions: quantum-limited sensitivity, extraordinary frequency coverage from MHz to THz, and complete immunity to antenna size constraints imposed by the Chu limit \cite{Anderson2021, Tu2024, Zhang2023d}.
Laboratory measurements confirm \gls{snr} gains exceeding $20$ dB in practical deployments \cite{Bussey2022}, with theoretical analyses projecting gains up to $44$ dB as quantum control techniques advance \cite{Chen2025}.
Recent experiments have demonstrated Mbps-scale data transmission over \gls{awgn} channels using standard modulation, including \gls{qam}, \gls{psk}, and \gls{pam} \cite{Du2022, Meyer2023}.
However, these implementations reveal a fundamental limitation: RA receivers can measure only the amplitude of \gls{rf} signals, inherently losing phase information critical for coherent detection \cite{Xu2025}.

This amplitude-only detection creates a non-linear transformation of the signal space, resulting in fundamental ambiguities for phase-dependent modulation schemes like \gls{qam} and \gls{psk} \cite{Liu2025a}.
Current RA receiver architectures employ a \gls{lo} to generate known reference signals, partially enabling phase recovery at the receiver side.
However, even with this reference signal, conventional modulation schemes including amplitude-based \gls{pam} remain inherently suboptimal for RA receivers \cite{Meyer2023}.
The fundamental challenge stems from the complex interaction between reference signals and channel fading coefficients, which transforms the detection geometry in ways that conventional modulations cannot properly accommodate.
This inherent mismatch between traditional modulation designs and RA receivers' unique detection characteristics necessitates a fundamentally new approach that explicitly accounts for the amplitude-only constraint, reference signal dynamics, and channel effects within a unified framework.

In this paper, a LO-aware adaptive modulation (LOAM) scheme is proposed, specifically designed for RA receivers that maximizes the amplitude difference between adjacent constellation points at the receiver.
The fundamental innovation lies in the co-linear arrangement of constellation points, with their orientation dynamically adapted to the phase difference between the reference signal and channel coefficient.
The LOAM scheme exhibits two distinct operational regimes determined by reference signal strength.
In the strong reference signal regime, constellation points are symmetrically distributed around the origin with constant spacing determined by the transmission power constraint.
Conversely, in the weak reference signal regime, the constellation adopts an asymmetric distribution where the spacing is dynamically optimized based on the interplay between power constraints, channel coefficients, and reference signal characteristics.
This paper rigorously derives the mathematical threshold that delineates these operational regimes.
Comprehensive numerical and simulation results validate that LOAM achieves remarkable performance improvements, outperforming conventional modulation schemes, including \gls{qam}, \gls{psk}, and \gls{pam}, by more than $45$ dB under typical operating conditions.

\section{Signal Model and Problem Formulation} \label{sec2}
This section first presents the operational principle of RA receivers through their physical architecture, then develops the corresponding signal model.
Based on this foundation, the fundamental challenges for signal modulation are identified, leading to the problem formulation.

\subsection{Operational Principle of RA Receivers}
\begin{figure}[t!]
	\centering
	\includegraphics[width=0.49\textwidth]{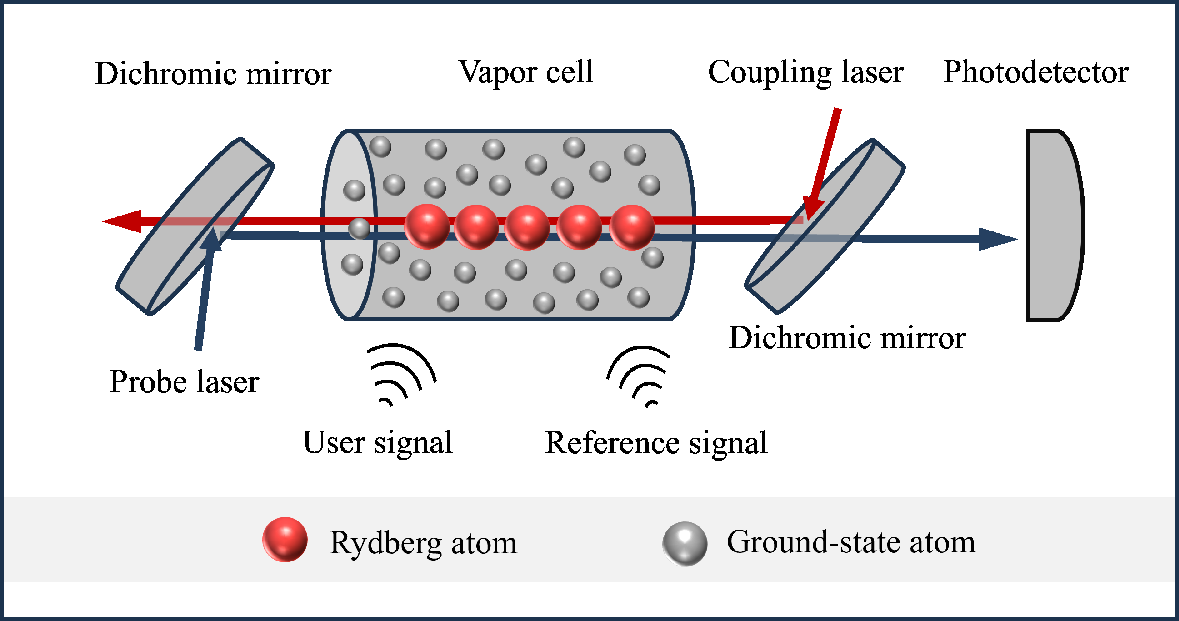} 
	\caption{Schematic of a RA receiver. The vapor cell contains ground-state atoms that are excited to Rydberg states via coupling and probe lasers. Incoming RF signals interact with these Rydberg atoms, modifying the transmission of the probe laser that is subsequently detected by the photodetector.}\label{receiver}
	\vspace{-1em}
\end{figure}

Fig. \ref{receiver} depicts the architecture of a typical RA receiver.
The core component is a vapor cell containing alkali atoms (such as rubidium or cesium) initially in their ground state.
Two laser systems (i.e., a probe laser and a coupling laser) establish a coherent superposition of atomic states via a three-level excitation scheme \cite{Cui2024}.
The coupling laser promotes atoms from an intermediate state to highly excited Rydberg states, distinguished by their large principal quantum numbers.
These Rydberg atoms possess extreme sensitivity to external electromagnetic fields owing to their substantial electric dipole moments.
When RF signals interact with the Rydberg atoms, they perturb the atomic energy levels, producing detectable variations in probe laser transmission.
A photodetector measures these transmission changes, which directly encode the amplitude of the incident RF signal.
Various detection mechanisms, including electromagnetically induced transparency (EIT), Autler-Townes splitting, and direct absorption measurements \cite{Cui2025}, all share the fundamental limitation of amplitude-only detection.

\subsection{Signal Model}
Based on the physical detection mechanism described above, the RA receiver effectively measures the amplitude of the combined signal consisting of the transmitted wireless signal and a LO-generated reference signal.
The mathematical model for this amplitude detection process is given by \cite{Cui2025}:
\begin{equation}
	z = |hx + b + n|,
\end{equation}
where $x \in \mathbb{C}$ denotes the transmitted symbol, $h \in \mathbb{C}$ represents the complex fading channel coefficient, $b \in \mathbb{C}$ is the LO-induced reference signal that is known and controllable at the receiver, $n \sim \mathcal{CN}(0, \sigma^2)$ denotes the additive white Gaussian noise (AWGN) with variance $\sigma^2$, and $z \in \mathbb{R}^+$ represents the detected amplitude at the receiver output.

The amplitude detection nature of RA receivers presents several challenges for signal modulation, making conventional constellation designs suboptimal.
First, the phase ambiguity occurs because the absolute value operation discards phase information.
Second, the reference signal creates a bias that significantly impacts the detection process and must be accounted for in modulation design.
Third, the non-linear transformation from amplitude detection creates a non-linear mapping of the signal space.

\subsection{Problem Formulation} \label{sec02b}
Given the transmitted symbol $x$ drawn from a finite alphabet set $\mathcal{X}$ of size $M > 1$ with equal probability, the optimal detection strategy employs maximum likelihood detection:
\begin{equation} \label{eqn11450426}
	\hat{x} = \underset{x \in \mathcal{X}}{\arg \min} \big|z - |hx + b| \big|^{2},
\end{equation}
which selects the constellation point $\hat{x}$ that minimizes the Euclidean distance between the received amplitude $z$ and the expected noiseless amplitude.
To facilitate analysis, define the transformed magnitude for each constellation point:
\begin{equation}
	r_{i} \triangleq |hx_{i} + b|,
\end{equation} 
where $i = 0, ..., M-1$, and $x_{i}$ denotes the $i$-th element in $\mathcal{X}$.
The detection strategy in \eqref{eqn11450426} reveals that optimal performance requires maximizing the minimum distance between transformed magnitudes. Thus, the constellation design problem is formulated as:
\begin{IEEEeqnarray} {rl}
	\mathcal{X}_{\text{opt}}\ &= \max_{\mathcal{X}} \Big\{\min_{i \neq j} |r_{i} - r_{j}|\Big\}, \label{eqn09570429} \\
	& \text{s.t. } \frac{1}{M}\sum_{i=1}^{M} |x_i|^{2} \leq P, \nonumber
\end{IEEEeqnarray}
where $P$ denotes the average power constraint.

This formulation explicitly demonstrates that optimal constellation design for RA receivers must account for both the LO-induced reference signal $b$ and the channel coefficient $h$.
Conventional modulation schemes fail to incorporate these critical factors, thereby achieving suboptimal performance.
This fundamental limitation motivates the development of the LOAM scheme, which adaptively adjusts constellation geometry based on these parameters to maximize detection performance.

\section{LOAM for RA Receivers}
This section develops the LOAM for RA receivers that dynamically adapts to both the LO-induced reference signal and channel coefficients.
The derivation of LOAM proceeds systematically through three fundamental insights: first establishing the co-linear structure of optimal constellation points, then proving the equal spacing property of transformed magnitudes, and finally deriving the precise mathematical expressions for the constellation points under different operational regimes.

The fundamental geometric structure (i.e., co-linearity) of the optimal constellation for RA receivers is characterized by the following result.

\begin{lem}[Co-linearity of the optimal constellation points] \label{lem01}
	Given known reference signal $b$ and channel coefficient $h$ ($h \neq 0$) at the transmitter, define
	\begin{equation}
		c \triangleq -b/h,
	\end{equation}
	Then, for any specified set of transformed magnitudes, the maximum power efficiency is achieved when all constellation points in $\mathcal{X}_{\text{opt}}$ lie co-linearly along the ray emanating from the origin through point $c$.
\end{lem}

\begin{IEEEproof}
	Given $c = -b/h$, the relationship $r_{i} = |hx_{i} + b|$ can be rewritten as
	\begin{equation} \label{eqn11330501}
		|x_i - c| = r_{i}/|h|. 
	\end{equation}
	Geometrically, equation \eqref{eqn11330501} describes a circle in the complex plane centered at $c$ with radius $r_{i}/|h|$. 
	The objective is to minimize $|x_{i}|$ (maximize power efficiency) while maintaining the constraint $r_{i} = |hx_{i} + b|$.
	Applying the triangle inequality yields:
	\begin{IEEEeqnarray}{ll}
		|x_{i}| = |(x_{i} - c) + c| 
		\geq \big||x_{i} - c| - |c|\big|. 
	\end{IEEEeqnarray}
	This lower bound is achieved if and only if vectors $(x_{i} - c)$ and $c$ are co-linear, requiring $x_{i}$ to lie on the line passing through both the origin and point $c$.
	Therefore, for any desired transformed magnitude $r_{i}$, the most power-efficient constellation point must lie on this specific ray.
\end{IEEEproof}

The result in \lemref{lem01} establishes that the optimal constellation for RA receivers must be one-dimensional, with all points co-linearly arranged along a specific ray determined by the channel coefficient $h$ and reference signal $b$.
This fundamental constraint arises from the amplitude-only detection mechanism, which renders conventional two-dimensional constellations such as \gls{qam} and \gls{psk} inherently suboptimal.
Furthermore, in fading channel environments where the phase of the channel coefficient varies dynamically, the optimal ray direction (through point $c = -b/h$) changes accordingly.
Existing one-dimensional modulation schemes like \gls{pam} fail to adapt their constellation direction to these variations, making them equally suboptimal for RA receivers.
Having established the necessity for an adaptive one-dimensional constellation structure, the subsequent analysis determines the optimal spacing between constellation points along the dynamically oriented ray.

\begin{lem}[Equal spacing of transformed magnitudes] \label{lem02}
	For a set of $M$ non-negative real numbers sorted as $\{r_{0} \leq ... \leq r_{M-1}\}$, the minimum pairwise distance $\delta = \min_{i \neq j}|r_{i} - r_{j}|$ is maximized when the magnitudes form an arithmetic progression:
	\begin{equation} \label{eqn09140429}
		r_{i} = r_{0} + i \delta,
	\end{equation}
	where $i = 0,...,M-1$.
\end{lem}

\begin{IEEEproof}
	For sorted magnitudes, the minimum pairwise distance equals the minimum consecutive difference:
	\begin{equation}
		\delta = \min_{i=1}^{M-1}(r_{i} - r_{i-1}).
	\end{equation}
	The total range can be decomposed as:
	\begin{equation}
		r_{M-1} - r_{0} = \sum_{i=1}^{M-1}(r_{i} - r_{i-1}).
	\end{equation}
	Since each consecutive difference satisfies $(r_{i} - r_{i-1}) \geq \delta$:
	\begin{equation}
		r_{M-1} - r_{0} \geq (M-1) \delta.
	\end{equation}
	This yields the upper bound:
	\begin{equation}
		\delta \leq \frac{r_{M-1} - r_{0}}{M-1}.
	\end{equation}
	This bound is achieved if and only if all consecutive differences equal $\delta$:
	\begin{equation}
		r_{i} - r_{i-1} = \delta, \quad \forall i.
	\end{equation}
	Therefore, maximum minimum distance requires equally spaced magnitudes, forming the arithmetic progression in \eqref{eqn09140429}.
\end{IEEEproof}

It is important to note that \lemref{lem02} establishes the equal spacing property purely from a distance optimization perspective, independent of any power constraints. 
The result holds for any set of non-negative real numbers $\{r_i\}$ within a fixed range $[r_0, r_{M-1}]$, regardless of how these magnitudes are generated or what constraints they satisfy at the transmitter.
This universality is crucial because it allows us to first optimize the spacing of transformed magnitudes to maximize detection performance, and then subsequently determine the corresponding constellation points that satisfy the power constraint.
The power constraint enters the problem only when mapping these optimally spaced magnitudes back to the constellation points $\{x_i\}$ through the relationship $r_i = |hx_i + b|$.
Consequently, both \lemref{lem01} and \lemref{lem02} establish fundamental geometric principles that transcend specific power regimes, forming the theoretical foundation for deriving power-constrained constellation points in subsequent analysis.

\begin{thm}\label{thm01}
	Given power constraint $P$ and constellation size $M$, the optimal constellation points $\mathcal{X}_{\text{opt}} = \{x_0, \dots, x_{M-1}\}$ with corresponding transformed magnitudes $\{r_{0} < \ldots < r_{M-1}\}$ are determined as:
	
	\textit{Case 1 (Strong reference signal):} When $|b|^2 \geq \frac{3P(M-1)|h|^2}{M+1}$,
	\begin{equation}
		x_{i} = e^{j\angle c} \left(- \frac{M-1}{2}d_{\text{max}} + id_{\text{max}} \right),
	\end{equation}
	where $d_{\text{max}}$ is given by \eqref{eqn10240502}.
	
	\textit{Case 2 (Weak reference signal):} When $|b|^2 < \frac{3P(M-1)|h|^2}{M+1}$,
	\begin{equation}
		x_{i} = e^{j\angle c} \big(|c| + id_{\text{anchor}}\big),
	\end{equation}
	where $d_{\text{anchor}}$ is given by \eqref{eqn10250502}.
\end{thm}

\begin{IEEEproof}
	See \appref{appthm01}.
\end{IEEEproof}


\thmref{thm01} reveals that strong reference signals yield larger spacing between transformed magnitudes than small reference signals, directly enhancing detection performance in RA receivers.
While the spacing parameter $d_{\text{max}}$ in the strong reference regime matches that of conventional \gls{pam}, LOAM achieves fundamental superiority through adaptive phase alignment.
The critical performance metric is the minimum distance between transformed magnitudes at the receiver.
LOAM with strong reference signals achieves:
\begin{equation}
	\delta_{\text{LOAM}} = |h|d_{\text{max}}.
\end{equation}
In contrast, conventional PAM suffers from phase misalignment.
Since PAM constrains its constellation to the real axis while the optimal direction is along the ray through point $c = -b/h$, geometric projection analysis yields:
\begin{equation}
	\delta_{\text{PAM}} = |h|d_{\text{max}}|\cos(\angle h - \angle b)|.
\end{equation}
This cosine factor represents the projection loss due to PAM's fixed orientation, making $\delta_{\text{PAM}} \leq \delta_{\text{LOAM}}$ with equality only when $\angle h = \angle b$.
Therefore, LOAM's adaptive alignment guarantees optimal detection performance across all channel conditions, while PAM remains fundamentally limited by its rigid structure.

Moreover, the proposed LOAM scheme naturally extends to the special case of zero reference signal ($|b| = 0$), yielding:
\begin{equation}
	x_{i} = id_{\text{LO-free}},
\end{equation}
where the optimal spacing becomes:
\begin{equation}
	d_{\text{LO-free}} = \sqrt{\dfrac{3P}{2(M-1)(2M-1)}}.
\end{equation}
In this LO-free scenario, LOAM maintains superiority over conventional schemes.
For example, in traditional modulation, antipodal symbols like $+1$ and $-1$ produce identical observations at LO-free RA receivers due to the amplitude-only detection, rendering them indistinguishable.
LOAM circumvents this fundamental limitation by employing a unipolar constellation structure optimized for amplitude detection.

\begin{figure*}[t]
	\centering
	\subfigure[LO-free]{
		\begin{minipage}[t]{0.32\textwidth}	
			\centering
			\includegraphics[width=5.8cm]{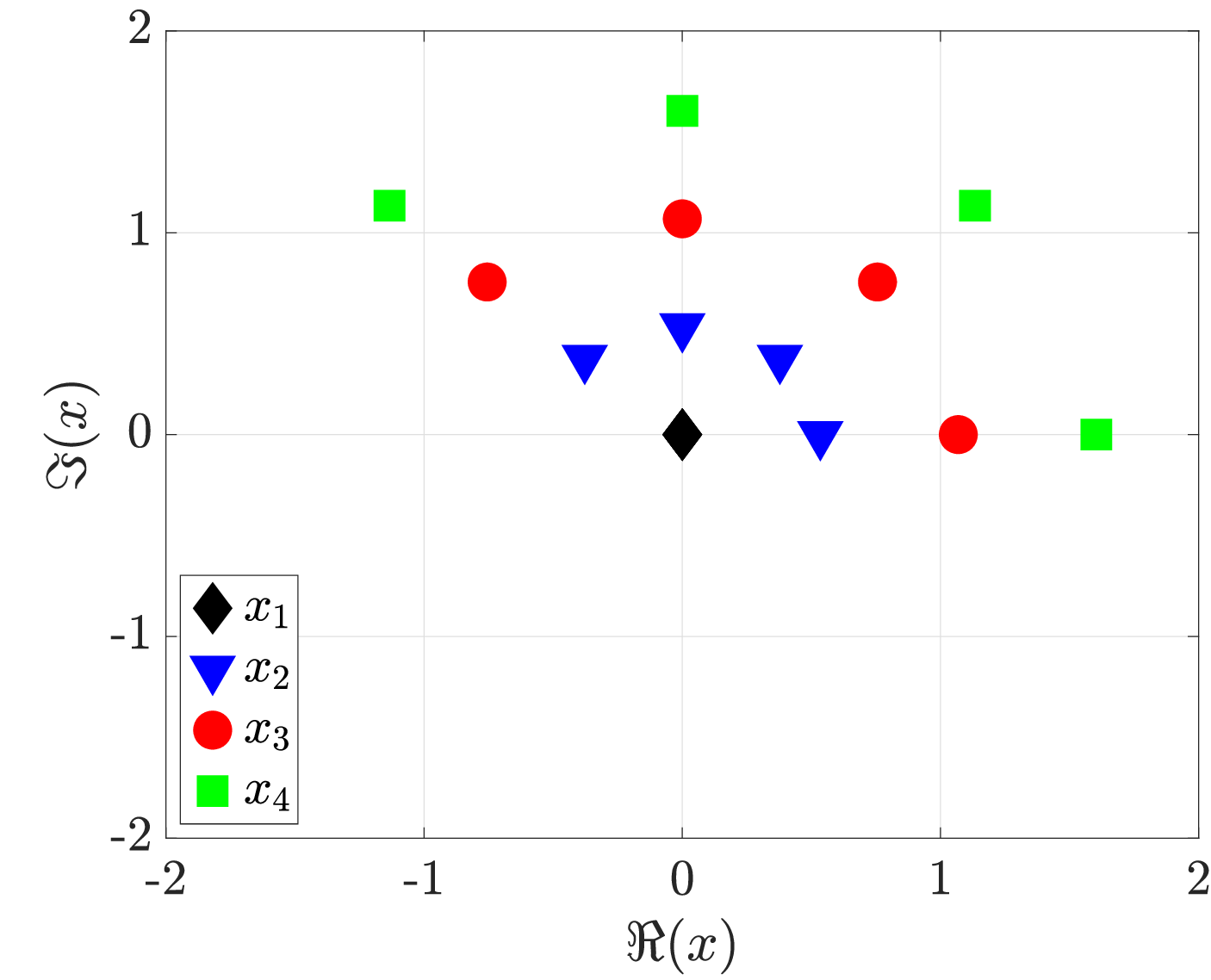}
	\end{minipage}}
	\subfigure[Weak reference signal]{
		\begin{minipage}[t]{0.32\textwidth}	
			\centering
			\includegraphics[width=5.8cm]{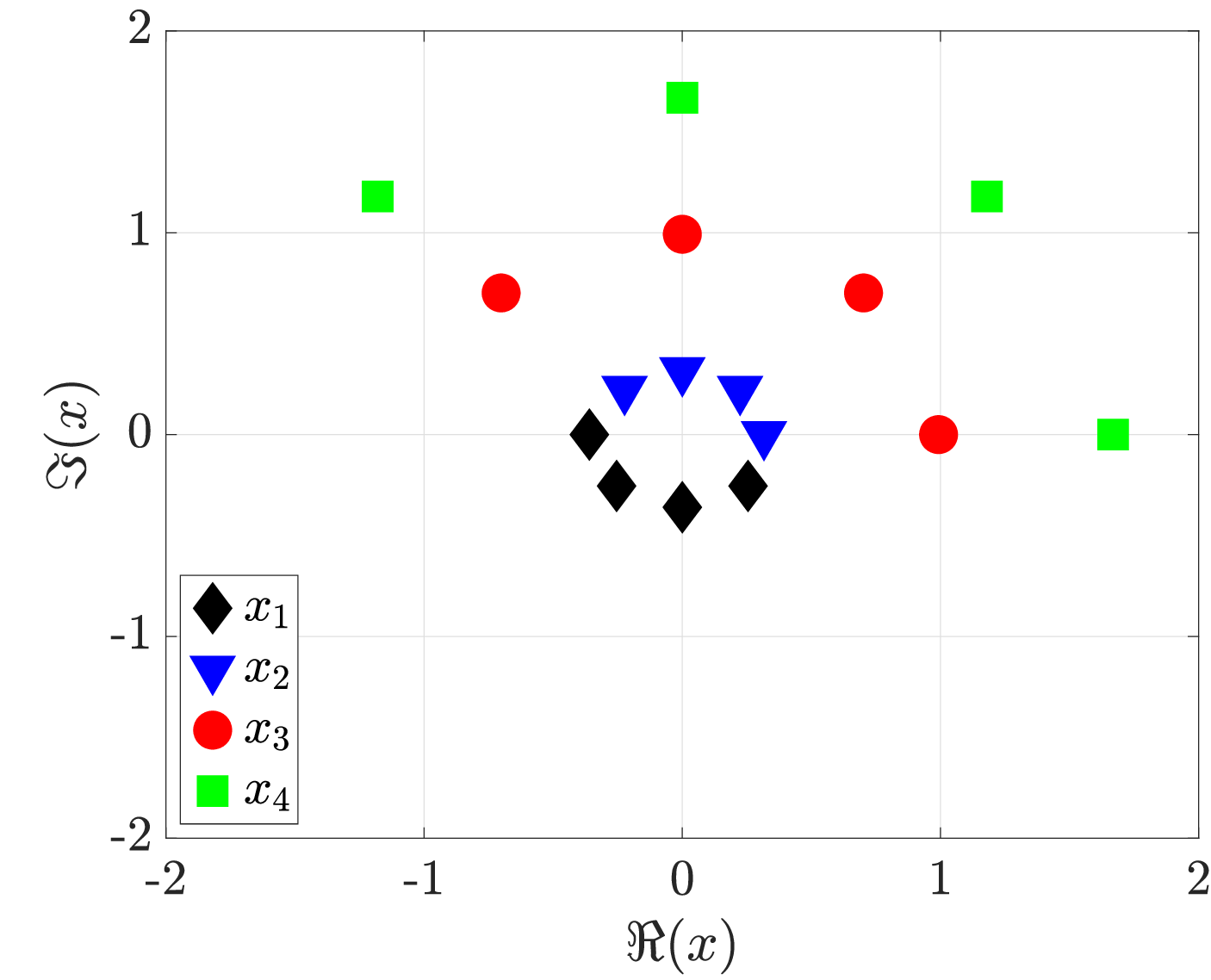}
	\end{minipage}}
	\subfigure[Strong reference signal]{
		\begin{minipage}[t]{0.32\textwidth}	
			\centering
			\includegraphics[width=5.8cm]{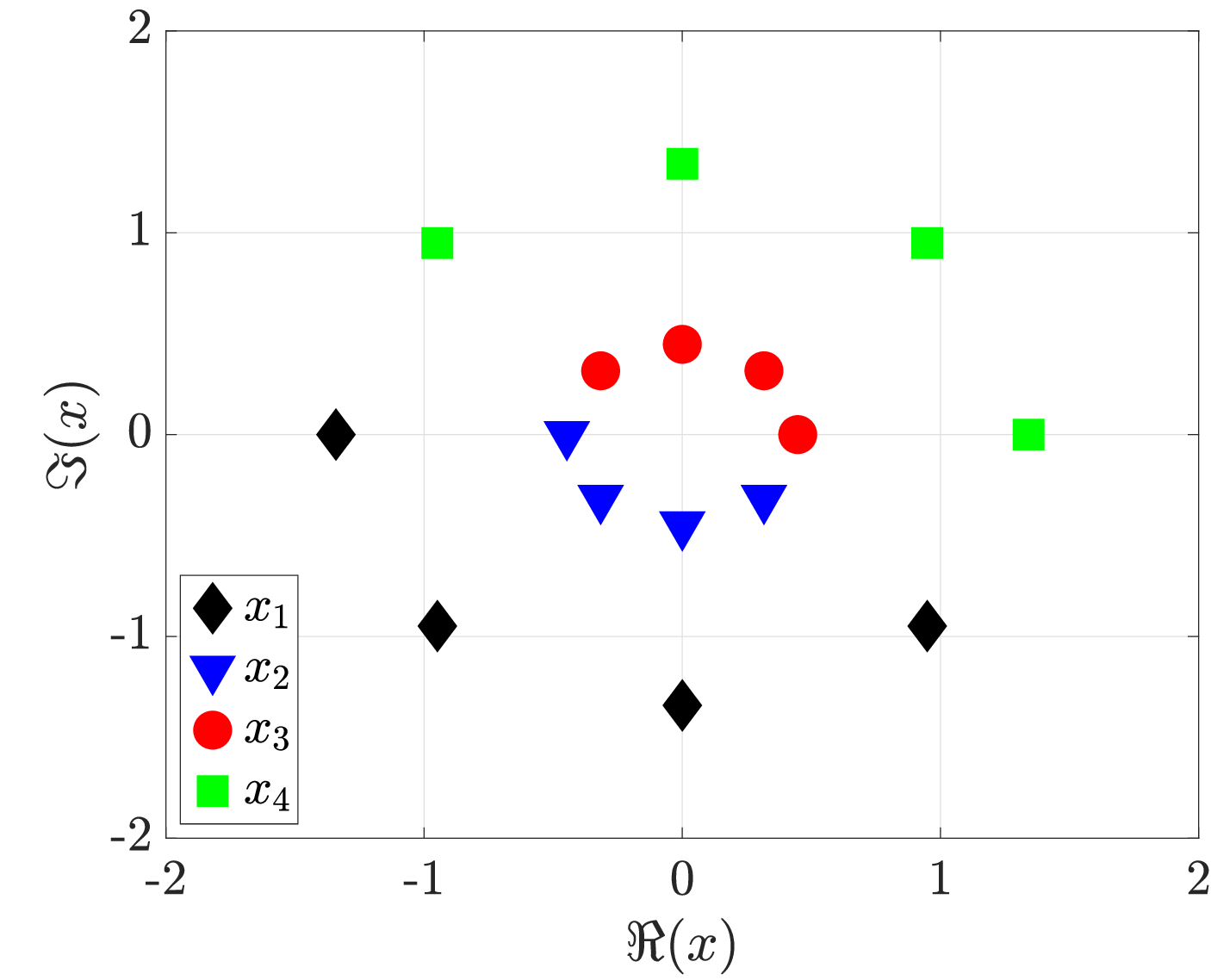}
	\end{minipage}}
	\caption{\label{fig:constellation}Visualization of LOAM constellation points with $M =4$ and the phase of $c$ is selected from the discrete set $\{0, \frac{\pi}{4}, \frac{\pi}{2}, \frac{3\pi}{4}\}$. (a) LO-free case showing constellation points co-linearly arranged along a ray from the origin. (b) Weak reference signal case displaying non-symmetrically distributed constellation points with the first point $x_{1}$ anchored at $c$. (c) Strong reference signal case showing symmetrically distributed constellation points centered around the origin.}
	\vspace{-1em}
\end{figure*}

\section{Numerical and Simulation Results} \label{secSimu}
This section validates the effectiveness of the proposed LOAM scheme through comprehensive numerical analysis and simulations.
The evaluation encompasses two primary objectives: visual demonstration of adaptive constellation patterns and quantitative performance comparisons with conventional modulation schemes across various reference signal conditions.
Throughout all experiments, we maintain consistent simulation parameters with power constraint $P=1$.
The following subsections present two case studies addressing these objectives.

\textit{Case Study 1 (Visualization of LOAM Constellation Geometry):}
Figure \ref{fig:constellation} illustrates the adaptive nature of LOAM by presenting three constellation diagrams for $M=4$ under different reference signal strengths, directly validating Theorem \ref{thm01}.
For the weak reference signal case, we set $|b|=\frac{P(M-1)|h|^2}{M+1}$ to clearly demonstrate the transition between operational regimes.
To comprehensively showcase the phase adaptation capability, we vary $\angle c \in \{0, \frac{\pi}{4}, \frac{\pi}{2}, \frac{3\pi}{4}\}$, revealing how LOAM dynamically aligns its constellation structure with channel conditions.

In subplot (a), the LO-free case ($|b| = 0$) is depicted. 
Here, all constellation points ($x_1$, $x_2$, $x_3$, $x_4$) are arranged co-linearly along a ray emanating from the origin.
The points follow an arithmetic progression with equal spacing, confirming \lemref{lem02}'s prediction about equal spacing being optimal.
This arrangement represents the special case where no reference signal is used, yet the constellation still maintains distinguishable transformed magnitudes at the receiver.

Subplot (b) illustrates the weak reference signal case.
The constellation exhibits a non-symmetric distribution around the origin.
Notably, the first point ($x_1$, black diamond) appears to be anchored at position $c$, while the remaining points ($x_2$, $x_3$, $x_4$) extend outward with equal spacing $d_{\text{anchor}}$.
This confirms the theoretical prediction from \thmref{thm01} that when the reference signal is not sufficiently strong, the optimal constellation has its first point at $c$ with subsequent points equally spaced along the ray.

Subplot (c) represents the strong reference signal case.
Here, the constellation points are symmetrically distributed around the origin, with equal spacing $d_{\text{max}}$ between consecutive points.
This symmetric arrangement allows for maximized minimum distance 
between transformed magnitudes, leading to better detection performance.
The black diamond points ($x_1$) now appear on both sides of the origin, visually confirming the symmetric nature of the optimal constellation when the reference signal is sufficiently strong.

The different spatial arrangements across the three scenarios demonstrate how the LOAM scheme adaptively configures the constellation based on reference signal strength to maximize detection performance.
The transition from a ray-based arrangement in the LO-free case to a symmetric distribution in the strong reference signal case illustrates how the scheme takes full advantage of stronger reference signals to improve spacing between transformed magnitudes.

\begin{figure*}[t]
	\centering
	\subfigure[LO-free; $M=4$]{
		\begin{minipage}[t]{0.32\textwidth}	
			\centering
			\includegraphics[width=5.8cm]{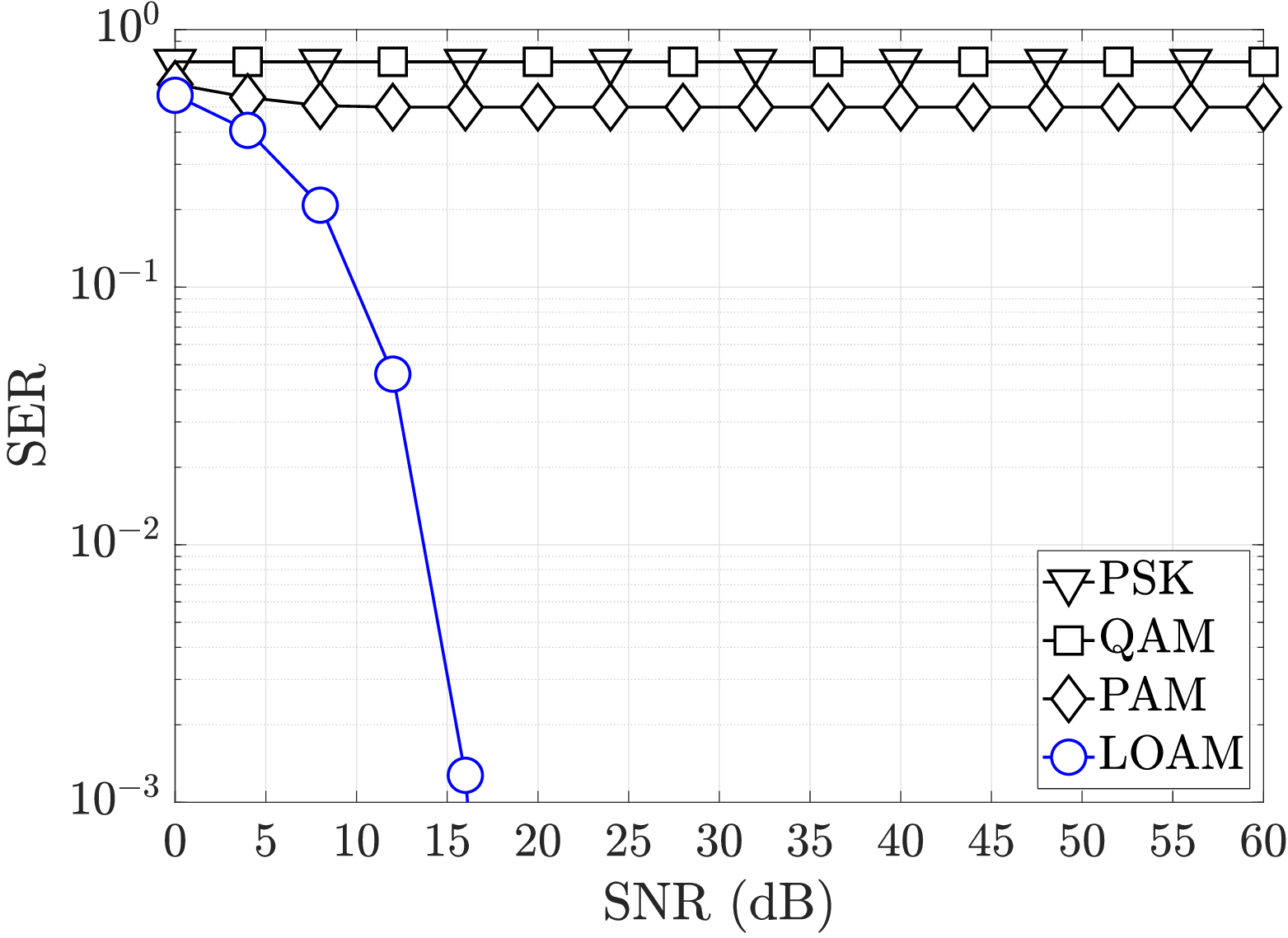}
	\end{minipage}}
	\subfigure[Weak reference signal; $M=4$]{
		\begin{minipage}[t]{0.32\textwidth}	
			\centering
			\includegraphics[width=5.8cm]{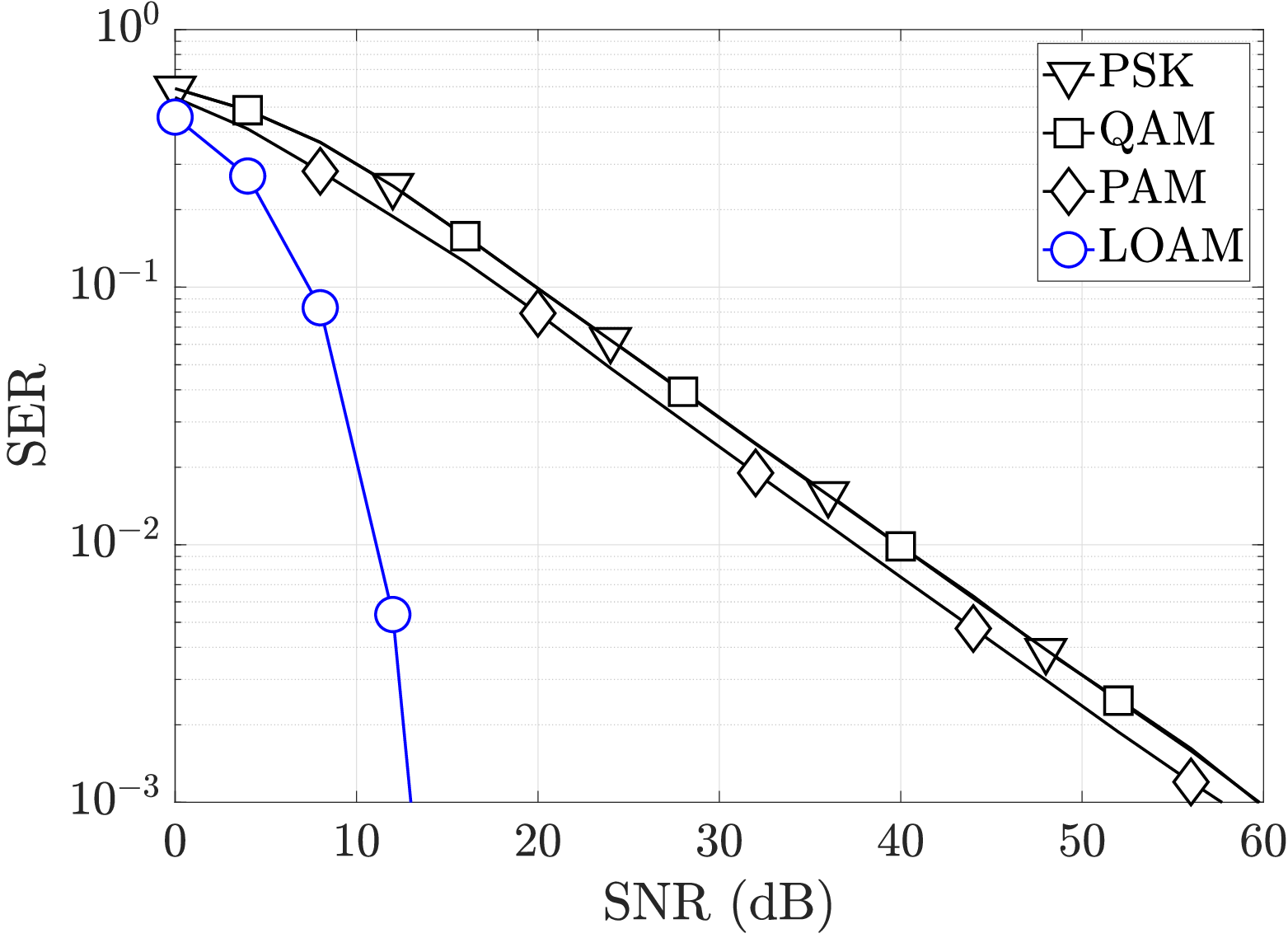}
	\end{minipage}}
	\subfigure[Strong reference signal; $M=4$]{
		\begin{minipage}[t]{0.32\textwidth}	
			\centering
			\includegraphics[width=5.8cm]{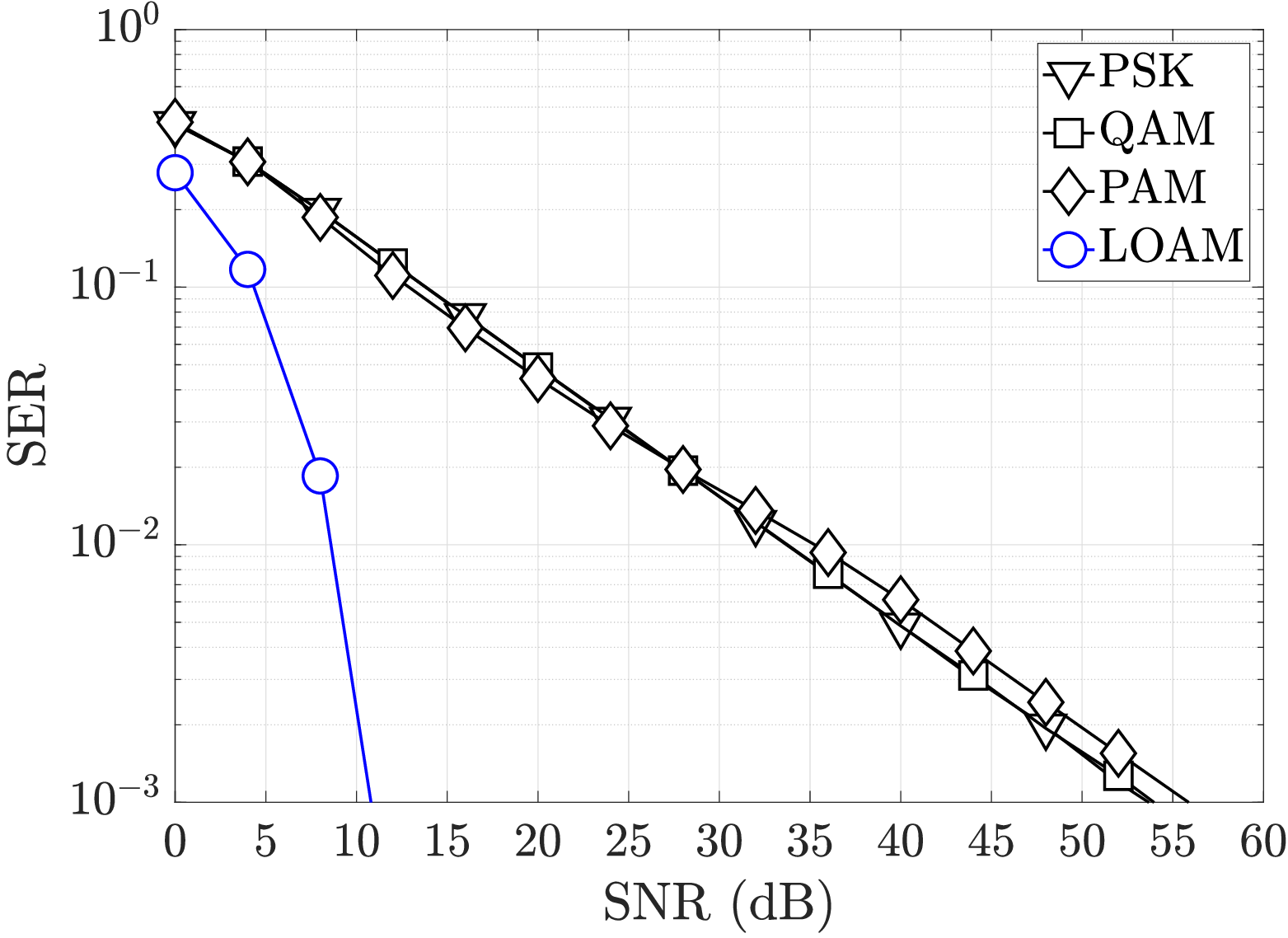}
	\end{minipage}}
	\subfigure[LO-free; $M=64$]{
		\begin{minipage}[t]{0.32\textwidth}	
			\centering
			\includegraphics[width=5.8cm]{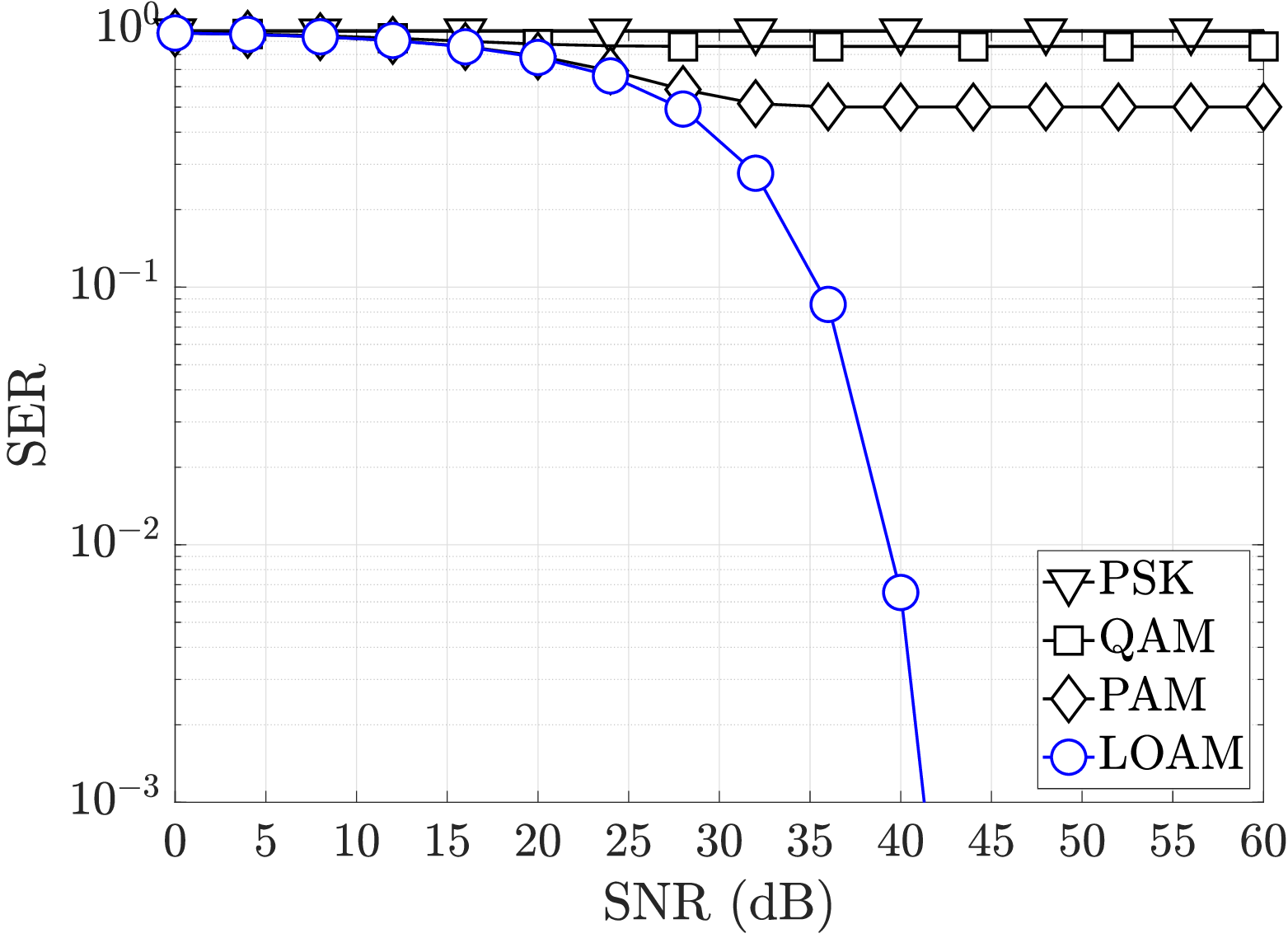}
	\end{minipage}}
	\subfigure[Weak reference signal; $M=64$]{
		\begin{minipage}[t]{0.32\textwidth}	
			\centering
			\includegraphics[width=5.8cm]{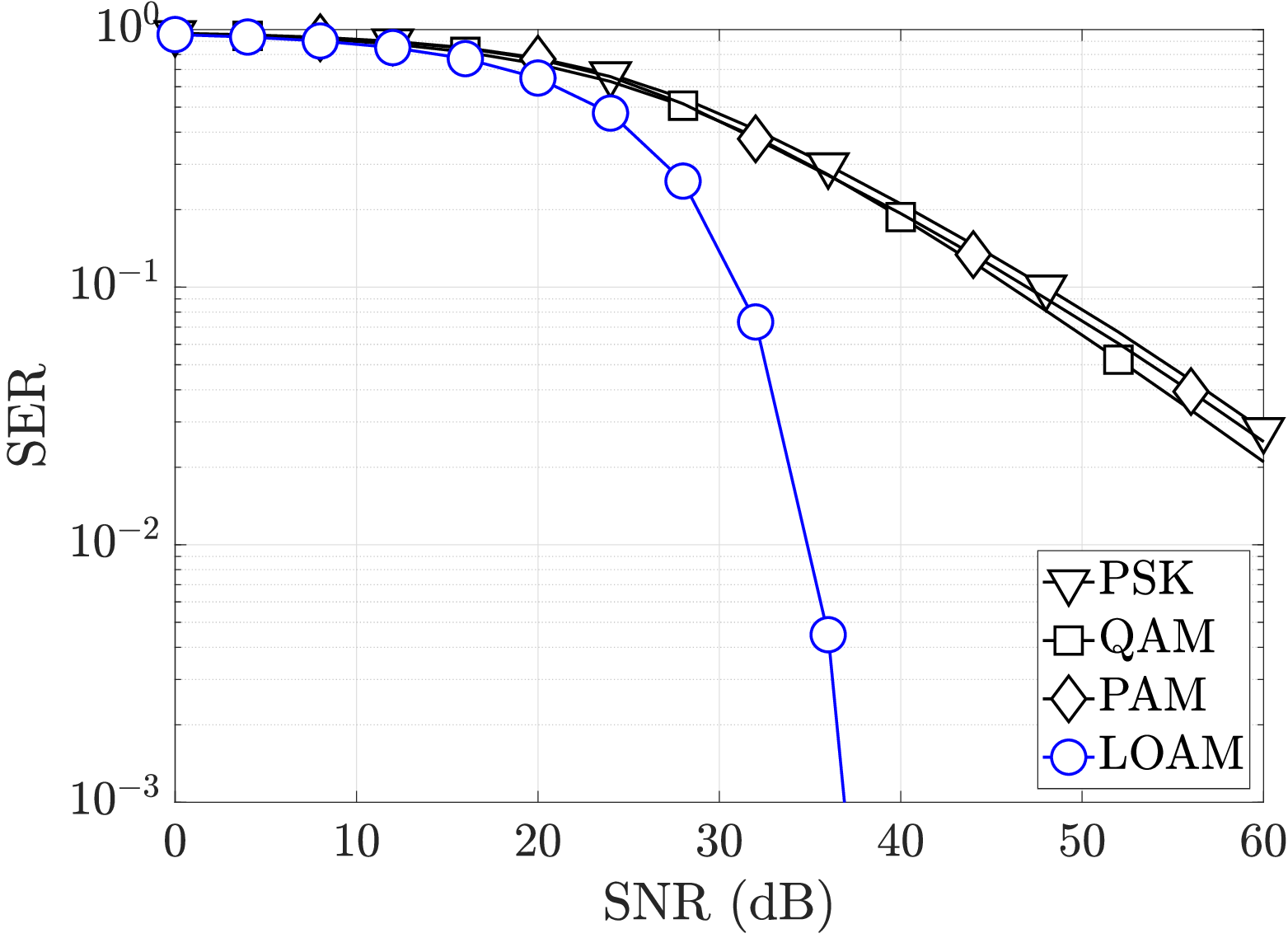}
	\end{minipage}}
	\subfigure[Strong reference signal; $M=64$]{
		\begin{minipage}[t]{0.32\textwidth}	
			\centering
			\includegraphics[width=5.8cm]{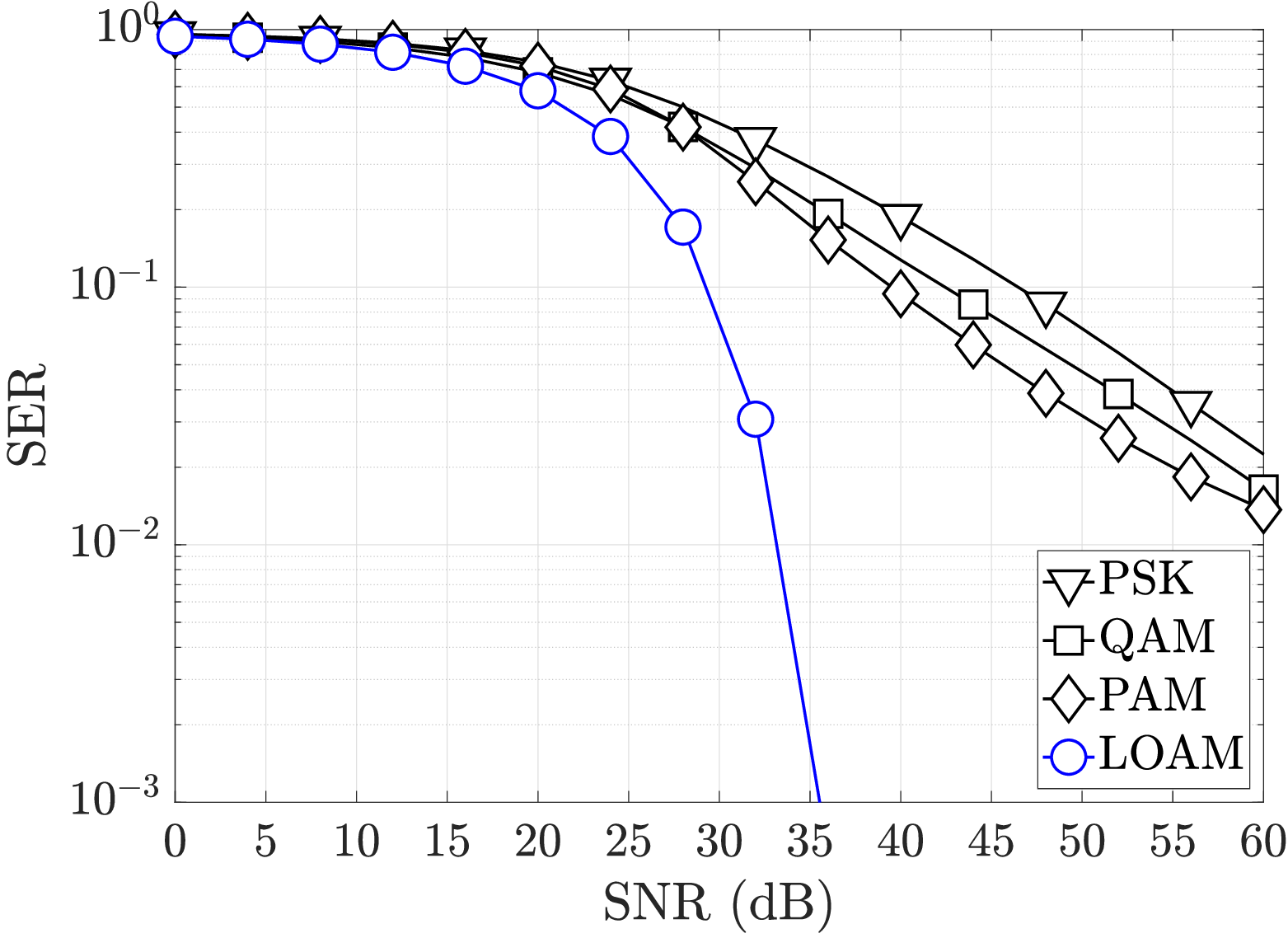}
	\end{minipage}}
	\caption{\label{fig03} SER performance comparison between the proposed LOAM scheme and conventional modulation schemes (PAM, QAM, PSK) under different reference signal conditions and constellation sizes.  Top row (a-c): $4$-point constellations ($M=4$). Bottom row (d-f): $64$-point constellations ($M=64$).}
	\vspace{-1em}
\end{figure*}

\textit{Case Study 2 (Comparison between LOAM and current modulation schemes):}
\figref{fig03} evaluates the \gls{ser} performance of LOAM against conventional modulation schemes (PAM, QAM, PSK) across various reference signal strengths and constellation sizes ($M = 4$ and $M = 64$).

For the LO-free scenario (Fig. \ref{fig03}a,d), LOAM demonstrates dramatic superiority over all conventional schemes. For $M = 4$, PAM, QAM, and PSK exhibit complete failure with SER plateauing at approximately $0.7$5, indicating random guessing among four symbols.
This validates our theoretical prediction that amplitude-only detection creates irresolvable ambiguities for conventional modulations.
In contrast, LOAM achieves progressive error reduction with increasing SNR, validating its optimized design for amplitude-only detection.
Under weak reference signals (Fig. \ref{fig03}b,e), all schemes show improved performance, but with distinct behaviors. For $M = 4$, LOAM maintains approximately $45$ dB advantage over conventional schemes at SER = $10^{-3}$. However, for $M = 64$, PSK emerges as competitive at high SNR, while LOAM and PAM exhibit comparable performance, reflecting the complex interplay between constellation density and reference signal strength.
With strong reference signals (Fig. \ref{fig03}c,f), the performance hierarchy becomes constellation-dependent.
For $M = 4$, LOAM retains its significant advantage.
For $M = 64$, PSK achieves the best performance, benefiting from the strong reference signal's ability to preserve phase information—a regime where PSK's circular symmetry becomes advantageous.

These results reveal several key insights that validate our theoretical framework. First, LOAM's superiority is most pronounced in challenging conditions, particularly LO-free or weak reference scenarios where conventional schemes struggle. Second, performance advantages scale with constellation size in amplitude-limited scenarios, demonstrating LOAM's robustness for higher-order modulations. Third, as reference signal strength increases, phase-based modulations gradually recover their effectiveness, explaining the observed performance crossovers. Fourth, these crossover points depend critically on both constellation size and reference signal strength, reflecting the complex interplay between detection mechanism and modulation structure. Finally, while conventional schemes exhibit high performance variability across different regimes, LOAM maintains consistent performance throughout.
These findings comprehensively validate the advantages of LOAM in all operating conditions, especially in amplitude-limited situations, which characterize actual RA receiver deployments.

\section{Conclusion}
This paper presented LOAM, a novel modulation scheme specifically engineered for RA receivers to overcome their inherent amplitude-only detection constraint.
Through rigorous theoretical analysis, we established that optimal performance requires co-linear constellation points arranged along an adaptive ray with equally spaced transformed magnitudes.
The derived constellation structure dynamically adjusts based on reference signal strength: symmetric distribution centered at the origin for strong reference signals, and asymmetric distribution anchored at point $c$ for weak reference signals.
Simulation results demonstrate that LOAM achieves remarkable performance gains exceeding $45$ dB over conventional modulation schemes (QAM, PSK, PAM) across all reference signal conditions.
Critically, LOAM maintains its superiority even in LO-free scenarios where traditional schemes fail completely due to phase ambiguity.
By fundamentally addressing the non-linear signal space transformation while preserving implementation simplicity, LOAM establishes a new paradigm for quantum-enhanced wireless communications.

\appendices

\section{Proof of \thmref{thm01}} \label{appthm01}
	For analytical simplicity, we rotate the complex plane to align $c$ with the real axis by defining:
\begin{equation}
	\tilde{c} \triangleq e^{-j\angle c} c = |c| \in \mathbb{R}^+,
\end{equation}
with corresponding constellation points $\tilde{x}_i = e^{-j\angle c} x_i$.
By \lemref{lem01}, all $\tilde{x}_i$ in $\mathcal{X}_{\text{opt}}$ are real-valued.
The transformed magnitude relationship becomes:
\begin{equation} \label{eqn10510502}
	r_i = |h| |\tilde{x}_i - \tilde{c}|.
\end{equation}
Since positioning any constellation point below $\tilde{c}$ would simultaneously reduce spacing and increase power consumption, optimality requires $\tilde{x}_{0} \geq \tilde{c}$, yielding:
\begin{equation}
	r_i = |h| (\tilde{x}_i - \tilde{c}), \quad \forall i.
\end{equation}
From \lemref{lem02}, the optimal spacing requires:
\begin{equation}
	\tilde{x}_i = \tilde{x}_0 + id,
\end{equation}
where $d = \frac{\delta}{|h|}$
The optimal constellation structure depends on the reference signal strength.

\textit{Case 1 (Strong reference signal regime):}
With sufficient reference signal strength, the power-optimal constellation centers at the origin:
\begin{equation} \label{eqn11430502}
	\tilde{x}_{i} = - \frac{M-1}{2}d + id.
\end{equation}
Note that $\tilde{x}_i^2 = x_i^2, \forall i$, the power constraint yields
\begin{equation}
	\frac{1}{M}\sum_{i=0}^{M-1} \tilde{x}_i^2 = \frac{(M^2-1)d^2}{12} \leq P,
\end{equation}
resulting in maximum spacing:
\begin{equation} \label{eqn10240502}
	d_{\text{max}} = \sqrt{\frac{12 P}{M^{2} - 1}}.
\end{equation}
The constraint $\tilde{x}_0 \geq \tilde{c}$ requires:
\begin{equation}
	-\frac{M-1}{2}d_{\text{max}} \geq \tilde{c},
\end{equation}
which holds when
\begin{equation}
	|b|^2 \geq \frac{3P(M-1)|h|^2}{M+1}.
\end{equation}

\textit{Case 2 (Weak reference signal regime):}
According to the discussion after \eqref{eqn10510502}, with insufficient reference signal strength, the optimal solution ``anchors'' the first point at $\tilde{c}$:
\begin{equation} \label{eqn11440502}
	\tilde{x}_{i} = \tilde{c} + id.
\end{equation}
Since $\tilde{x}_i^2 = x_i^2, \forall i$, the power constraint yields
\begin{equation}
	\frac{1}{M}\sum_{i=0}^{M-1} \tilde{x}_{i}^2 = \frac{1}{M}\sum_{i=0}^{M-1} (\tilde{c} + id)^2 \leq P.
\end{equation}
Solving this quadratic inequality yields:
\begin{equation} \label{eqn10250502}
	d_{\text{anchor}} = \dfrac{-3\tilde{c}}{4M-2} + \sqrt{\dfrac{3\tilde{c}^{2}}{2(2M-1)^{2}} - \dfrac{3(\tilde{c}^{2}-P)}{2(M-1)(2M-1)}}.
\end{equation}
Finally, we restore the original complex plane orientation, as follows
\begin{equation}
	x_i = e^{j\angle c} \tilde{x}_i,
\end{equation}
where $\tilde{x}_i$ is given by \eqref{eqn11430502} for Case 1 and \eqref{eqn11440502} for Case 2.
This completes the proof.

\section*{Acknowledgment}
This work was funded by the 5G and 6G Innovation Centre, University of Surrey.

\ifCLASSOPTIONcaptionsoff
\newpage
\fi

\bibliographystyle{IEEEtran}
\bibliography{../IEEEabrv,../thesis_list}
\end{document}

%% file: Acronyms.tex
\newacronym{1d}{1D}{one-dimensional}
\newacronym{2d}{2D}{two-dimensional }
\newacronym{3d}{3D}{three-dimensional}
\newacronym{3gpp}{3GPP}{3rd Generation Partnership Project}
\newacronym{5g}{5G}{fifth-generation}
\newacronym{6g}{6G}{sixth-generation}

\newacronym{ap}{AP}{access point}
\newacronym{amp}{AMP}{approximate message passing}
\newacronym{ask}{ASK}{amplitude shift keying}
\newacronym{awgn}{AWGN}{additive white Gaussian noise}

\newacronym{ber}{BER}{bit error rate}
\newacronym{bfgs}{BFGS}{Broyden–Fletcher–Goldfarb–Shanno}
\newacronym{bler}{BLER}{bit error rate}
\newacronym{bp}{BP}{belief propagation}
\newacronym{bpsk}{BPSK}{binary phase shift keying}
\newacronym{bs}{BS}{base station}

\newacronym{cbsm}{CBSM}{correlation-based stochastic model}
\newacronym{csirs}{CSI-RS}{channel state information reference signal}
\newacronym{cdf}{CDF}{cumulative distribution function}
\newacronym{cdma}{CDMA}{code division multiple access}
\newacronym{cg}{CG}{conjugate gradient descent}
\newacronym{clt}{CLT}{central limit theorem}
\newacronym{csi}{CSI}{Channel state information }
\newacronym{cvp}{CVP}{closest vector problem}

\newacronym{dof}{DoF}{degrees of freedom}
\newacronym{dsqt}{DSQT}{dual-slot quadrature transmission}

\newacronym{edw}{EDW}{exponentially decaying window}
\newacronym{elaa}{ELAA}{extremely large aperture array}
\newacronym{etsi}{ETSI}{European Telecommunications Standards Institute}

\newacronym{ff}{FF}{far-field}
\newacronym{fcsd}{FCSD}{fixed-complexity sphere decoder}
\newacronym{fec}{FEC}{forward error correction}
\newacronym{fspl}{FSPL}{free space path loss}

\newacronym{gbsm}{GBSM}{geometry-based stochastic model}
\newacronym{gd}{GD}{gradient descent}
\newacronym{gmsk}{GMSK}{Gaussian minimum shift keying}
\newacronym{gs}{GS}{Gauss-Seidel}
\newacronym{gsm}{GSM}{global system for mobile communication}

\newacronym{iid}{i.i.d.}{independently and identical distributed}
\newacronym{ils}{ILS}{integer least-squares}
\newacronym{ind}{i.n.d.}{independently and non-identical distributed}
\newacronym{imt}{IMT}{International Mobile Telecommunications}
\newacronym{isac}{ISAC}{integrated sensing and communication}
\newacronym{isi}{ISI}{intersymbol interference}
\newacronym{itur}{ITU-R}{International Telecommunication Union Radiocommunication Sector}
\newacronym{iui}{IUI}{inter-user interference}

\newacronym{ji}{JI}{Jacobi iteration}
\newacronym{jsac}{JSAC}{joint sensing and communication}

\newacronym{las}{LAS}{likelihood ascent search}
\newacronym{lbfgs}{LBFGS}{limited-memory Broyden–Fletcher–Goldfarb–Shanno}
\newacronym{leo}{LEO}{low Earth orbit}
\newacronym{lll}{LLL}{Lenstra-Lenstra-Lov\'{a}sz}
\newacronym{llr}{LLR}{log-likelihood ratio}
\newacronym{los}{LoS}{line of sight}
\newacronym{lo}{LO}{local oscillator}
\newacronym{lr}{LR}{lattice reduction}
\newacronym{lmmse}{LMMSE}{minimum mean square error}
\newacronym{ls}{LS}{least square}
\newacronym{lsd}{LSD}{list sphere decoder}
\newacronym{lte}{LTE}{long-term evolution}

\newacronym{map}{MAP}{maximum a posteriori}
\newacronym{mf}{MF}{matched filter}
\newacronym{mimo}{MIMO}{multiple-input multiple-output}
\newacronym{mld}{MLD}{maximum likelihood detection}
\newacronym{ml}{ML}{maximum likelihood}
\newacronym{mmimo}{mMIMO}{massive multiple-input multiple-output}
\newacronym{mrc}{MRC}{maximum ratio combining}
\newacronym{mse}{MSE}{mean square error}
\newacronym{ms}{MS}{matrix-splitting}
\newacronym{mt}{MT}{mobile terminal}

\newacronym{nf}{NF}{near-field}
\newacronym{nlos}{NLoS}{non-LoS}

\newacronym{od}{OD}{orthogonality defect}

\newacronym{pam}{PAM}{pulse-amplitude modulation}
\newacronym{pdf}{PDF}{probability distribution function}
\newacronym{pda}{PDA}{probabilistic data association}
\newacronym{pep}{PEP}{pairwise error probability}
\newacronym{pl}{PL}{path-loss}
\newacronym{pmf}{PMF}{probability mass function}
\newacronym{psk}{PSK}{phase-shift keying}
\newacronym{pwm}{PWM}{plane-wave model}

\newacronym{qam}{QAM}{quadrature amplitude modulation}
\newacronym{qpsk}{QPSK}{quadrature phase shift keying}
\newacronym{qn}{QN}{quasi-Newton}

\newacronym{rts}{RTS}{reactive tabu search}
\newacronym{ri}{RI}{Richardson iteration}
\newacronym{ris}{RIS}{reconfigurable intelligent surface}
\newacronym{rf}{RF}{radio-frequency}
\newacronym{rpe}{RPE}{rotational phase encoding}
\newacronym{rsr}{RSR}{reference-to-signal ratio}
\newacronym{rss}{RSS}{received signal strength}
\newacronym{rzf}{RZF}{regularized-ZF}

\newacronym{sa}{SA}{Seysen's algorithm}
\newacronym{sd}{SD}{steepest descent}
\newacronym{sdr}{SDR}{semidefinite relaxation}
\newacronym{ser}{SER}{symbol error rate}
\newacronym{sf}{SF}{shadow fading}
\newacronym{sic}{SIC}{succesive interference cancellation}
\newacronym{sinr}{SINR}{signal-to-interference-plus-noise ratio}
\newacronym{siso}{SISO}{single input single output}
\newacronym{snr}{SNR}{signal-to-noise ratio}
\newacronym{sns}{SNS}{spatial non-stationarity}
\newacronym{sota}{SoTA}{state-of-the-art}
\newacronym{ssor}{SSOR}{symmetric successive over-relaxation}
\newacronym{svd}{SVD}{singular value decomposition }
\newacronym{swm}{SWM}{spherical-wave model}

\newacronym{tr}{TR}{Technical Report}
\newacronym{ts}{TS}{tabu search}

\newacronym{uca}{UCA}{uniform cylindrical array}
\newacronym{ue}{UE}{user equipment}
\newacronym{ula}{ULA}{uniform linear array}
\newacronym{ura}{URA}{uniform rectangular array}
\newacronym{uma}{UMa}{urban macro}
\newacronym{umi}{UMi}{urban micro}
\newacronym{upa}{UPA}{uniform planar array}
\newacronym{ut}{UT}{user terminal}

\newacronym{vblast}{V-BLAST}{vertical Bell Labs layered space-time}

\newacronym{xlmimo}{XL-MIMO}{extra-large multiple-input multiple-output}

\newacronym{zf}{ZF}{zero-forcing}